\documentclass[a4paper,12pt]{article}

\begin{document}
\vspace{5 pt}

\centerline{\large \bf The second "Einstein paradox".}

\vspace{7 pt}
\centerline{\sl V.A.Kuz'menko}
\vspace{5 pt}
\centerline{\small \it Troitsk Institute of Innovation and Fusion Research,}
\centerline{\small \it Troitsk, Moscow region, 142190, Russian Federation.}
\vspace{5 pt}
\begin{abstract}

	Time invariance problem of a photon absorption 
process in atoms and molecules is discussed.

\vspace{5 pt}
\end{abstract}

\vspace{12 pt}

The Dirac equation, in the case of electromagnetic interaction, is not 
invariant under unitary time inversion [1]. However, a common opinion exists 
that the electromagnetic interaction in nature must be time invariant [2]. 
The attempts to find what is a proof  basis of such point of view in a 
special case of a photon absorption process in atoms and molecules give only 
indistinct references point at Einstein's works [3]. 

If one keeps in mind the Einstein coefficients of absorption and stimulated 
emission, this basis is erroneous. The Einstein coefficients characterize 
integral cross-section of optical transition.  Time invariance preserving 
demands equality not only the integral cross-section, but also it demands 
equality of spectral width of forward and backward optical transitions. 
Einstein nothing writes about the width of optical transitions. So, 
Einstein coefficients have no direct connection to T-invariance of photon 
absorption process.

There is also exists the Einstein's opinion "that physics could be 
restricted to the time-symmetric case for which retarded and advanced fields 
are equivalent" [4]. Obviously there is a main basis of existing point of 
view, because of any experimental result in proof of T-invariance preserving 
in a photon absorption process is absent. 

In contrast, for the opposite point of view we have one direct and complete 
experimental proof and a number of indirect evidences. The direct and 
complete experimental proof is connected with the experimental study of the 
so-called line wings [5]. The experiments clearly show a very strong 
T-invariance violation in a photon absorption process in molecules. Although 
the integral cross-section of forward and backward optical transitions are 
obviously the same, the spectral width and cross-section for such transitions 
differ on several order of magnitude [6]. 

The concept of T-invariance violation of a photon absorption process is a 
good basis for explanation of most effects in nonlinear optics from a pure 
quantum position without using any semiclassical approximation [7]. There are 
indirect experimental proofs. The most striking example is the population 
transfer effect in the case of sweeping a resonance conditions in a two level 
system. On the basis of T-invariance violation of absorption process this  
effect has a simple and natural explanation [8]. In contrast, on the basis of 
semiclassical wave approximation theory the explanation of this effect looks
\begin{minipage}[t]{138mm}
like as the ravings of a madman [9].

\hspace*{\parindent}  On the whole such situation may be 
called as the second "Einstein paradox"\footnote[1]{The first paradox is 
connected with quantum statistics [10,11].}. When great authority and 
delusion of one scientist delay on decades progress of physical theory in 
some fields of scientific research. So, do we have "Einstein paradox" in 
quantum optics?
\end{minipage}


\vspace{5 pt}

\begin{thebibliography}{99}
\bibitem{1} W.M.Jin, eprint, quant-ph/0001029
\bibitem{2} V.B.Berestetsky, E.M.Lifshits, L.P.Pitaevsky, {\sl"Relativistic 
Quantum Theory"}, part 1, Nauka, Moscow, p.66 (1968) in Russian.
\bibitem{3} A.K.Popov, eprint, quant-ph/0005118
\bibitem{4} A.Bohm, N.L.Harshman, eprint, quant-ph/9805063, p.2
\bibitem{5} V.A.Kuz'menko, eprint, aps1998dec29\_002,  \\
http://publish.aps.org/eprint/
\bibitem{6} V.A.Kuz'menko, eprint, hep-ph/0002084
\bibitem{7} V.A.Kuz'menko, eprint, physics/0007076
\bibitem{8} V.A.Kuz'menko, eprint, aps1999sep30\_002, \\
http://publish.aps.org/eprint/
\bibitem{9} R.L.Shoemaker, in {\sl"Laser and Coherence Spectroscopy"}, 
Ed. J.I.Steinfeld, N.Y., Plenum, p.197 (1978).
\bibitem{10} Ya.M.Gel'fer, V.L.Lyuboshits, M.I.Podgoretsky, 
{\sl"Gibbs Paradox and Identity of Particles in Quantum Mechanics"}, Nauka, 
Moscow (1975), in Russian.
\bibitem{11} P.Ehrenfest, G.E.Uhlenbeck, {\sl"Paul Ehrenfest"}, Collected 
Scientific Papers, North-Holland Publishing Company, Amsterdam, p.539, (1959).
\end{thebibliography}
\end{document}